\documentclass[twocolumn,aps,prc,superscriptaddress,showpacs,floatfix]{revtex4}
\usepackage{amssymb}
\usepackage{graphicx}

\begin{document}

\title{Pseudorapidity dependence of anisotropic flows in relativistic
heavy-ion collisions}
\author{Lie-Wen Chen}
\affiliation{Department of Physics, Shanghai Jiao Tong University, Shanghai 200030, China}
\affiliation{Cyclotron Institute and Physics Department, Texas A\&M University, College
Station, Texas 77843-3366}
\author{Vincenzo Greco}
\affiliation{Cyclotron Institute and Physics Department, Texas A\&M University, College
Station, Texas 77843-3366}
\author{Che Ming Ko}
\affiliation{Cyclotron Institute and Physics Department, Texas A\&M University, College
Station, Texas 77843-3366}
\author{Peter F. Kolb}
\affiliation{Physik Department, Technische Universit\"{a}t M\"{u}nchen, D-85747 Garching,
Germany}
\date{\today }

\begin{abstract}
The pseudorapidity dependence of anisotropic flows $v_{1}$, $v_{2}$, $v_{3}$%
, and $v_{4}$ of charged hadrons in heavy-ion collisions at the Relativistic
Heavy Ion Collider is studied in a multi-phase transport model. We find that
while the string melting scenario, in which hadrons that are expected to be
formed from initial strings are converted to their valence quarks and
antiquarks, can explain the measured $p_{T}$-dependence of $v_{2}$ and $%
v_{4} $ of charged hadrons at midrapidity with a parton scattering cross
section of about $10$ \textrm{mb}, the scenario without string melting
reproduces better the recent data on $v_{1}$ and $v_{2}$ of charged hadrons
at large pseudorapidity in Au + Au collisions at $\sqrt{s}=200$ \textrm{AGeV}%
. Our results thus suggest that a partonic matter is formed during early
stage of relativistic heavy ion collisions only around midrapidity and that
strings remain dominant at large rapidities. The $p_{T}$-dependence of $%
v_{1} $, $v_{2}$, $v_{3}$ and $v_{4}$ for charged hadrons at forward
pseudorapidity is also predicted, and we find that while $v_{1}$ and $v_{2}$
are appreciable at large pseudorapidity the higher-order anisotropic flows $%
v_{3}$ and $v_{4} $ are essentially zero.
\end{abstract}

\maketitle

\section{Introduction}

Anisotropic flows in heavy ion collisions \cite{Ollit92,Rqmd,Danie98,Zheng99}%
, which have been studied extensively at the Relativistic Heavy Ion Collider
(\textrm{RHIC}), are sensitive to the properties of produced matter. This
sensitivity not only exists in the larger elliptic flow \cite%
{tean,kolb,huov,Zhang99,moln1,Lin:2001zk,gyulv2,Voloshin03} but also in the
smaller higher-order anisotropic flows \cite%
{Kolb99,Teaney99,Kolb00,Kolb03,STAR03,chen04,Kolb04}. Furthermore, scaling
relations among hadron anisotropic flows are observed in the experimental
data \cite{STAR03}, and they are shown in theoretical models to relate to
similar scaling relations among parton elliptic flows \cite{chen04,Kolb04}.
These studies have, however, mainly focused on hadrons at mid-rapidity where
odd-order anisotropic flows vanish in collisions of equal mass nuclei.
Recently, anisotropic flows at finite pseudorapidities have also been
measured in Au + Au collisions at \textrm{RHIC}. The experimental results
show that both $v_{1}$ and $v_{2}$ depend strongly on the rapidity \cite%
{STAR03,phobos02,manly03,tonjes04,oldenburg04,back04}, and this has so far
not been reproduced by theoretical models \cite{hirano01,heinz04,csanad04}.
These new experimental data thus offer the opportunity to test the validity
of theoretical models and to study the dynamics and properties of produced
matter at large rapidity in heavy ion collisions at \textrm{RHIC}.

In the present work, we shall use a multi-phase transport (\textrm{AMPT})
model, that includes both initial partonic and final hadronic interactions 
\cite{Zhang:2000bd,Lin:2001cx}, to study the pseudorapidity dependence of
anisotropic flows $v_{1}$, $v_{2}$, $v_{3}$, and $v_{4}$ of charged hadrons
in heavy-ion collisions at \textrm{RHIC}. Both the default version and the
version with string melting, i.e., allowing hadrons that are expected to be
formed from initial strings to convert to their valence quarks and
antiquarks \cite{Lin:2001zk,LinHBT02,ko}, will be used. The latter was able
to explain the measured $p_{T}$ dependence of $v_{2}$ and $v_{4}$ of
mid-rapidity charged hadrons with a parton scattering cross section of about 
$10$ \textrm{mb}. We find in the present study that the same parton cross
section fails to reproduce recent data on $v_{1}$ and $v_{2}$ at large
pseudorapidity in Au + Au collisions at $\sqrt{s}=200$ \textrm{AGeV}. These
data are explained instead by the default scenario without string melting.
Our results thus suggest that a pure partonic matter is formed only near
midrapidity during the early stage of these collisions, and the matter at
large pseudorapidity remains dominated by strings. We also give predictions
on the $p_{T}$-dependence of $v_{1}$, $v_{2}$, $v_{3}$ and $v_{4}$ at
forward pseudorapidity.

\section{The AMPT model}

The \textrm{AMPT} model \cite{Zhang:2000bd,Lin:2001cx,zhang,pal} is a hybrid
model that uses minijet partons from hard processes and strings from soft
processes in the Heavy Ion Jet Interaction Generator (\textrm{HIJING}) model 
\cite{Wang:1991ht} as the initial conditions for modeling heavy ion
collisions at ultra-relativistic energies. Time evolution of resulting
minijet partons is then described by Zhang's parton cascade (\textrm{ZPC})%
\cite{Zhang:1997ej} model. At present, this model includes only
parton-parton elastic scatterings with an in-medium cross section given by: 
\begin{equation}  \label{crscp}
\frac{d\sigma _{p}}{dt}=\frac{9\pi \alpha _{s}^{2}}{2}\left( 1+{\frac{{\mu
^{2}}}{s}}\right) \frac{1}{(t-\mu ^{2})^{2}},
\end{equation}%
where the strong coupling constant $\alpha _{s}$ is taken to be $0.47$, and $%
s$ and $t$ are usual Mandelstam variables. The effective screening mass $\mu 
$ depends on the temperature and density of the partonic matter but is taken
as a parameter in \textrm{ZPC} for fixing the magnitude and angular
distribution of parton scattering cross section. After minijet partons stop
interacting, they are combined with their parent strings, as in the \textrm{%
HIJING} model with jet quenching, to fragment into hadrons using the Lund
string fragmentation model as implemented in the \textrm{PYTHIA} program 
\cite{Sjostrand:1994yb}. The final-state hadronic scatterings are then
modeled by a relativistic transport (\textrm{ART}) model \cite{Li:1995pr}.
The default \textrm{AMPT} model \cite{Zhang:2000bd} has been quite
successful in describing measured rapidity distributions of charge
particles, particle to antiparticle ratios, and spectra of low transverse
momentum pions and kaons \cite{Lin:2001cx} in heavy ion collisions at the
Super Proton Synchrotron (\textrm{SPS}) and \textrm{RHIC}. It has also been
useful in understanding the production of $J/\psi $ \cite{zhang} and
multistrange baryons \cite{pal} in these collisions.

Since the initial energy density in Au + Au collisions at \textrm{RHIC} is
much larger than the critical energy density at which the hadronic matter to
quark-gluon plasma transition would occur \cite{Kharzeev:2001ph,zhang}, the 
\textrm{AMPT} model has been extended to convert the initial excited strings
into partons \cite{Lin:2001zk}. In this string melting scenario, hadrons
(mostly pions), that would have been produced from string fragmentation, are
converted instead to valence quarks and/or antiquarks with current quark
masses. Interactions among these partons are again described by the\textrm{\
ZPC} parton cascade model. Since there are no inelastic scatterings, only
quarks and antiquarks from the melted strings are present in the partonic
matter. The transition from the partonic matter to the hadronic matter is
then achieved using a simple coalescence model, which combines two nearest
quark and antiquark into mesons and three nearest quarks or antiquarks into
baryons or anti-baryons that are close to the invariant mass of these
partons. The present coalescence model is thus somewhat different from the
ones recently used extensively \cite{greco,hwa,fries,molnar03} for studying
hadron production at intermediate transverse momenta. Using parton
scattering cross sections of $6$-$10$ \textrm{mb}, the \textrm{AMPT} model
with string melting is able to reproduce both the centrality and transverse
momentum (below $2$ \textrm{GeV}$/c$) dependence of the elliptic flow \cite%
{Lin:2001zk} and pion interferometry \cite{LinHBT02} measured in Au+Au
collisions at $\sqrt{s}=130$ \textrm{AGeV} at \textrm{RHIC} \cite%
{Ackermann:2000tr,STARhbt01}. It has also been used for studying the kaon
interferometry in these collisions \cite{lin}. We note that the above cross
sections are significantly smaller than that needed to reproduce the parton
elliptic flow from the hydrodynamic model \cite{molnar}. The resulting
hadron elliptic flows in the \textrm{AMPT} model with string melting are,
however, amplified by modeling hadronization via quark coalescence \cite%
{molnar03}, leading to a satisfactory reproduction of experimental data.

\section{Transverse momentum spectra}

\begin{figure}[th]
\includegraphics[scale=0.89]{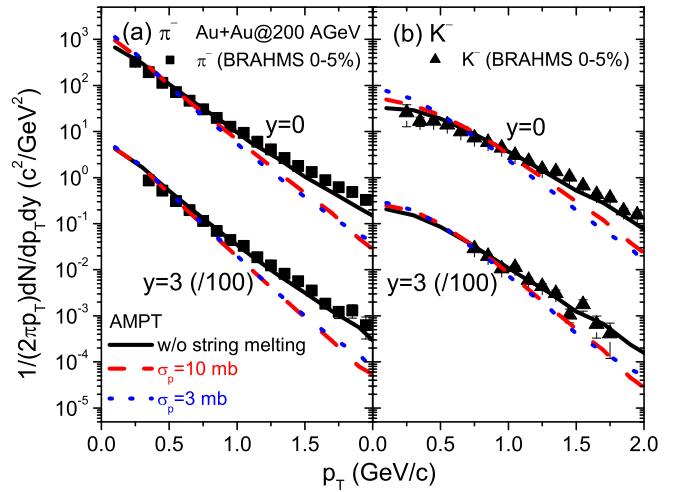}
\caption{{\protect\small (Color online) Pion and kaon transverse momentum
spectra at rapidity }$y=0${\protect\small \ and }$3${\protect\small \ in
central Au+Au collisions at }$\protect\sqrt{s}=200${\protect\small \ AGeV.
Solid curves are from default AMPT model, while dotted and dashed curves are
from AMPT model with string melting using parton cross section }$\protect%
\sigma _{p}=3${\protect\small \ mb and 10 mb, respectively. Experimental
data from the BRAHMS collaboration \protect\cite{brahms} are squares for
pions and triangles for kaons.}}
\label{pty}
\end{figure}

We first show in Fig.\ref{pty} the pion and kaon transverse momentum spectra
at rapidity $y=0$ and $3$ in central Au+Au collisions at $\sqrt{s}=200$
AGeV. Compared to experimental data from the BRAHMS collaboration \cite%
{brahms}, shown by squares for pions and triangles for kaons, the default 
\textrm{AMPT} model shown by solid curves reproduces them well, but the 
\textrm{AMPT} model with string melting, on the other hand, gives a smaller
inverse slope parameter for both parton cross sections of $\sigma _{p}=3$ mb
(dotted curves) and $10$ mb (dashed curves). The reason that hadron
transverse momentum spectra are softened in the string melting scenario is
due to a softer initial parton spectra obtained from converting hadrons to
quarks and antiquarks, and the small current quark masses that make their
transverse momentum spectra less affected by radial collective flow than
hadrons in the default \textrm{AMPT} model. Since hadron anisotropic flows
are given by ratios of hadron momentum distributions in the transverse
plane, the \textrm{AMPT} model with string melting is expected to give a
reliable prediction \cite{Lin:2001zk}.

\section{Anisotropic flows}

The anisotropic flows $v_{n}$ of particles are the Fourier coefficients in
the decomposition of their transverse momentum spectra in the azimuthal
angle $\phi $ with respect to the reaction plane \cite{Posk98}, i.e., 
\begin{equation}
E\frac{d^{3}N}{dp^{3}}=\frac{1}{2\pi }\frac{dN}{p_{T}dp_{T}dy}%
[1+\sum_{n=1}^{\infty }2v_{n}(p_{T},y)\cos (n\phi )]  \label{dndphi}
\end{equation}%
Because of the symmetry $\phi \leftrightarrow -\phi $ in the collision
geometry, no sine terms appear in the above expansion. For particles at
midrapidity in collisions with equal mass nuclei, anisotropic flows of odd
orders vanish as a result of the additional symmetry $\phi \leftrightarrow
\phi +\pi $. The anisotropic flows generally depend on particle transverse
momentum and rapidity, and for a given rapidity the anisotropic flows at
transverse momentum $p_{T}$ can be evaluated according to 
\begin{equation}
v_{n}(p_{T})=\left\langle \cos (n\phi )\right\rangle ,  \label{vn1}
\end{equation}%
where $\left\langle \cdot \cdot \cdot \right\rangle $ denotes average over
the azimuthal distribution of particles with transverse momentum $p_{T}$.
The anisotropic flows $v_{n}$ can further be expressed in terms of the
single-particle averages: 
\begin{eqnarray}
v_{1}(p_{T}) &=&\left\langle \frac{p_{x}}{p_{T}}\right\rangle  \label{v1} \\
v_{2}(p_{T}) &=&\left\langle \frac{p_{x}^{2}-p_{y}^{2}}{p_{T}^{2}}%
\right\rangle  \label{v2} \\
v_{3}(p_{T}) &=&\left\langle \frac{p_{x}^{3}-3p_{x}p_{y}^{2}}{p_{T}^{3}}%
\right\rangle  \label{v3} \\
v_{4}(p_{T}) &=&\left\langle \frac{p_{x}^{4}-6p_{x}^{2}p_{y}^{2}+p_{y}^{4}}{%
p_{T}^{4}}\right\rangle  \label{v4}
\end{eqnarray}%
where $p_{x}$ and $p_{y}$ are, respectively, the projections of particle
momentum in and perpendicular to the reaction plane.

\section{Pseudorapidity dependence of anisotropic flows}

\begin{figure}[th]
\includegraphics[scale=0.9]{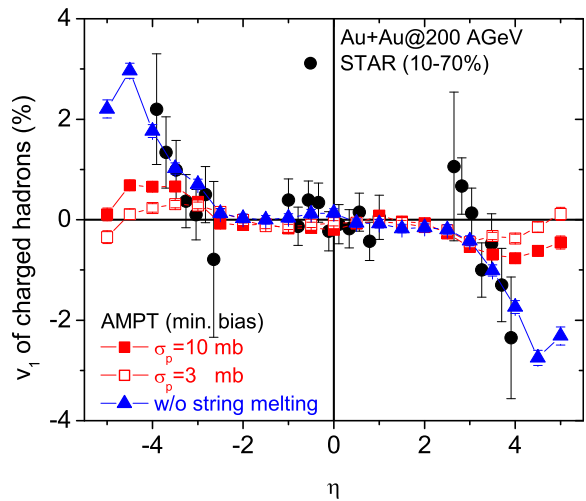}
\caption{{\protect\small (Color online) Pseudorapidity dependence of }$v_{1}$%
{\protect\small \ from minimum bias events of Au + Au collisions at }$%
\protect\sqrt{s}=200${\protect\small \ AGeV in the string melting scenario
with parton scattering cross sections }$\protect\sigma _{p}=3$%
{\protect\small \ (open squares) and }$10${\protect\small \ (solid squares)
\ mb as well as the scenario without string melting (triangles). Data are
from the STAR collaboration (circles) \protect\cite{STAR03}.}}
\label{v1yh}
\end{figure}

In Fig. \ref{v1yh}, we show the pseudorapidity dependence of $v_{1}$ for
charged hadrons from minimum bias events of Au + Au collisions at $\sqrt{s}%
=200$ \textrm{AGeV} by using the string melting scenario with parton
scattering cross sections $\sigma _{p}=3$ (open squares) and $10$ \textrm{mb}
(solid squares) and the scenario without string melting (default \textrm{AMPT%
} model, triangles). Also included in Fig. \ref{v1yh} are recent data from
the STAR collaboration (circles) \cite{STAR03}. Both scenarios can reproduce
approximately the data around the mid-pseudorapidity region, i.e., $v_{1}$
is flat (essentially zero) around mid-$\eta $. For $v_{1}$ at large $%
\left\vert \eta \right\vert $, the string melting scenario with both parton
scattering cross sections $\sigma _{p}=3$ \textrm{mb} and $10$ \textrm{mb}\
underestimates significantly the data. On the other hand, the scenario
without string melting seems to give a good description of $v_{1}$ at large $%
\left\vert \eta \right\vert $. Our results thus indicate that the matter
produced at large $\left\vert \eta \right\vert $ ($\left\vert \eta
\right\vert \geq 3$) at \textrm{RHIC} initially consists of mostly strings
instead of partons. This is a reasonable picture as particles at large
rapidity are produced later in time when the volume of the system is large
and the energy density is small.

\begin{figure}[th]
\includegraphics[scale=0.9]{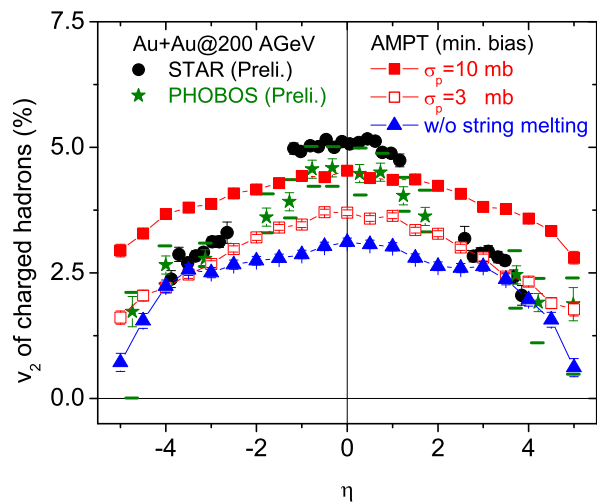}
\caption{{\protect\small (Color online) Pseudorapidity dependence of }$v_{2}$%
{\protect\small \ from minimum bias events of Au + Au collisions at }$%
\protect\sqrt{s}=200${\protect\small \ AGeV in the string melting scenario
with parton scattering cross sections }$\protect\sigma _{p}=3$%
{\protect\small \ (open squares) and }$10${\protect\small \ (solid squares)
\ mb as well as the scenario without string melting (triangles). Data are
from the PHOBOS (stars) \protect\cite{manly03} and STAR collaborations
(circles) \protect\cite{oldenburg04}.}}
\label{v2yh}
\end{figure}

The predicted pseudorapidity dependence of charged hadron $v_{2}$ from the
same reaction is shown in Fig. \ref{v2yh}, together with preliminary data
from the PHOBOS collaboration (stars) \cite{manly03} and the STAR
collaboration (circles) \cite{oldenburg04}. One sees that the string melting
scenario with $\sigma _{p}=$ $10$ \textrm{mb} (solid squared) describes very
well the data on $v_{2}$ around mid-$\eta $ ($\left| \eta \right|\leq 1.5$)
while it overestimates the data at large pseudorapidity. Surprisingly, the
calculated results with $\sigma _{p}=$ $10$ \textrm{mb} are similar to the
prediction from the hydrodynamic model that includes the ``thermalization
coefficient'' correction \cite{heinz04}. The overestimation of $v_{2}$ for
charged hadrons at large pseudorapidity ($\left| \eta \right| \geqslant 1.5$%
) obtained with $\sigma _{p}=$ $10$ \textrm{mb} may be due to the constant
parton scattering cross section used in the calculations. As shown in Eq.(%
\ref{crscp}), the parton scattering cross section depends on the gluon
screening mass and is thus temperature and density dependent. Since the
dynamics of partonic matter at different rapidities may not be the same in
heavy ion collisions at \textrm{RHIC}, different parton cross sections may
have to be used. Comparison between theoretical results and the experimental
data on elliptic flow indicates that a larger $\sigma _{p}=$ $10$ \textrm{mb}
is needed at midrapidity but a smaller $\sigma _{p}=$ $3$ \textrm{mb} (open
squares) gives a better description at large pseudorapidity. Also shown in
Fig. \ref{v2yh} are results obtained from the scenario without string
melting (triangles), and they are seen to also describe the data at large
pseudorapidity ($\left| \eta \right|\geqslant 3$). Therefore, the scenario
without string melting can describe simultaneously the data for $v_{1}$ and $%
v_{2}$ at large pseudorapidity ($\left| \eta \right| \geqslant 3$). These
interesting features imply that initially the matter produced at large
pseudorapidity ($\left| \eta \right| \geqslant 3$) is dominated by strings
while that produced around mid-rapidity ($\left| \eta \right| \leq 3$)
mainly consists of partons.

\section{$p_{T}$-dependence of anisotropic flows at forward rapidity}

\label{momentum}

\begin{figure}[th]
\includegraphics[scale=0.9]{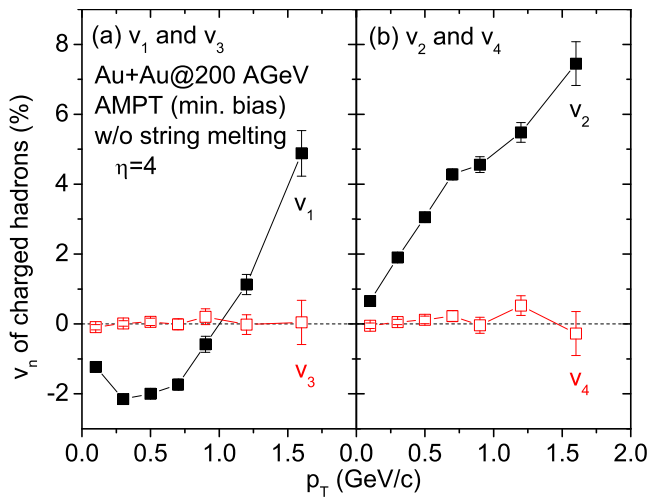}
\caption{{\protect\small (Color online)} {\protect\small Transverse momentum
dependence of }$v_{1}${\protect\small \ and }$v_{3}${\protect\small \ (a) as
well as }$v_{2}${\protect\small \ and }$v_{4}${\protect\small \ (b)\ for
charged hadrons at forward pseudorapidity (}$\protect\eta =4${\protect\small %
) from minimum bias events of Au + Au collisions at }$\protect\sqrt{s}=200$%
{\protect\small \ AGeV from the default AMPT model without string melting.}}
\label{vnpte4}
\end{figure}

More detailed information about anisotropic flows can be obtained from the
differential anisotropic flows, i.e., their $p_{T}$ dependence. Using the
scenario without string melting, we show in Fig. \ref{vnpte4} the $p_{T}$%
-dependence of $v_{1}$, $v_{2}$, $v_{3}$, and $v_{4}$\ for charged hadrons
at forward pseudorapidity ($\eta =4$) from minimum bias events of Au + Au
collisions at $\sqrt{s}=200$ \textrm{AGeV}. It is seen that the directed
flow $v_{1}(p_{T})$ shown in Fig. \ref{vnpte4} (a) is non-zero and changes
from negative to positive values at a balance transverse momentum of about $%
1.0$ \textrm{GeV}/\textsl{c}. This feature implies that charged hadrons with
lower and higher transverse momentum move preferentially towards the
negative and positive transverse flow direction, respectively, consistent
with that seen in the hydrodynamic model \cite{heinz04}. The $p_{T}$%
-integrated directed flow $v_{1}$ can be non-zero at large $\eta $ as
already shown in Fig. \ref{v1yh}. The differential anisotropic flow,
however, allows one to learn in more detail the dynamics of heavy ion
collisions. The $v_{3}(p_{T})$ shown in Fig. \ref{vnpte4} (a) is essentially
zero. Fig. \ref{vnpte4} (b) shows the differential elliptic flow $%
v_{2}(p_{T})$ and $v_{4}(p_{T})$ at $\eta =4$. A strong elliptic flow $%
v_{2}(p_{T})$ is observed while $v_{4}(p_{T})$ is again almost zero (less
than $0.5\%$), which is consistent with preliminary data from the STAR
collaboration \cite{oldenburg04}. Comparison of our predictions with future
data can test the conclusion that the matter produced at large
pseudorapidity ($\left| \eta \right| \geqslant 3$) in the early stage of
collisions is dominated by strings. In this case, it would be interesting to
see experimentally if the constituent quark number scaling of meson and
baryon elliptic flows seen at midrapidity is indeed not valid at large
pseudorapidity.

\section{Summary}

In summary, using the \textrm{AMPT} model, we have studied the
pseudorapidity dependence of anisotropic flows $v_{1}$, $v_{2}$, $v_{3}$,
and $v_{4}$ of charged hadrons in heavy-ion collisions at \textrm{RHIC}.
Within the string melting scenario, we find that a parton scattering cross
section of about $10$ \textrm{mb}, that is used in explaining the measured $%
p_{T}$-dependence of $v_{2}$ and $v_{4}$\ for charged hadrons at
midrapidity, fails to reproduce recent data on their $v_{1}$ and $v_{2}$ at
large pseudorapidity from Au + Au collisions at $\sqrt{s}=200$ \textrm{AGeV}%
. Allowing a smaller parton cross section at large pseudorapidity, the
measured pseudorapidity dependence of $v_{2}$ could be quantitatively
accounted for but the $v_{1}$ at large pseudorapidity ($\left| \eta \right|
\geqslant 3$) is still underestimated. We further find that $v_{1}$ and $%
v_{2}$ at large pseudorapidity can be described simultaneously by the
scenario without string melting. Our results thus suggest that the matter
produced at large pseudorapidity ($\left| \eta \right| \geqslant 3$) during
the early stage of Au + Au collisions at $\sqrt{s}=200$ \textrm{AGeV }is
dominated by strings while that produced around mid-pseudorapidity ($\left|
\eta \right| \leq 3$) consists mainly of partons. The predicted $p_{T}$%
-dependence of $v_{1}$, $v_{2}$, $v_{3}$ and $v_{4}$ at forward
pseudorapidity shows that there exist strong directed flow $v_{1}$ and
elliptic flow $v_{2}$ while $v_{3}$ and $v_{4}$ are essentially zero.
Experimental verification of these predictions at \textrm{RHIC} will be
useful in testing the \textrm{AMPT} model and in understanding how the
collision dynamics changes with rapidity.

\begin{acknowledgments}
This paper was based on work supported in part by the US National Science
Foundation under Grant No. PHY-0098805 and the Welch Foundation under Grant
No. A-1358 (LWC,VG,CMK), the National Science Foundation of China under
Grant No. 10105008 (LWC), the National Institute of Nuclear Physics (INFN)
in Italy (VG), and the BMBF and the DFG (PK).
\end{acknowledgments}

\end{document}